\documentclass[superscriptaddress,twocolumn,showpacs,amsmath,amssymb]{revtex4}
\usepackage{graphicx}
\usepackage{color}

\renewcommand{\vec}[1]{\mbox{\boldmath $#1$}}

\begin{document}

\title{
Investigating multi-channel quantum tunneling in heavy-ion fusion 
reactions with Bayesian spectral deconvolution}

\author{K. Hagino}
\affiliation{
Department of Physics, Tohoku University, Sendai 980-8578,  Japan}
\affiliation{Research Center for Electron Photon Science, Tohoku
University, 1-2-1 Mikamine, Sendai 982-0826, Japan}
\affiliation{
National Astronomical Observatory of Japan, 2-21-1 Osawa,
Mitaka, Tokyo 181-8588, Japan}


\begin{abstract}
Excitations of colliding nuclei during a nuclear reaction considerably 
affect fusion cross sections at energies around the Coulomb barrier. 
It has been demonstrated that such channel coupling effects can 
be represented in terms of a distribution of multiple fusion barriers. 
I here apply a Bayesian approach to analyze the so called fusion barrier 
distributions. This method 
determines simultaneously 
the barrier parameters and the number of 
barriers. 
I particularly investigate the $^{16}$O+$^{144}$Sm and 
$^{16}$O+$^{154}$Sm systems in order to demonstrate 
the effectiveness of the method. 
The present analysis indicates that the fusion barrier distribution 
for the former system is most consistent with three fusion barriers, 
even though the experimental data show only two distinct peaks. 
\end{abstract}

\pacs{
25.70.Jj, 
24.10.Eq, 
02.50.-r  
} 

\maketitle

Quantum tunneling plays an important role in many areas of 
physics and chemistry. One good example is a nuclear fusion 
reaction in stars, which is responsible for stellar energy 
production and nucleosynthesis \cite{Adelberger98}. 
While one considers a penetration of a one-dimensional barrier in many 
applications, it is important to notice that in reality 
quantum tunneling often involves many 
degrees of freedom \cite{Caldeira81} and/or takes place in a multi-dimensional 
space \cite{TN94,SBDN09}. 
It is naturally expected that this results in 
a significant modification in tunneling 
rates. 

In nuclear physics, a heavy-ion fusion reaction at energies around the 
Coulomb barrier is a typical example of quantum tunneling with many 
degrees of freedom \cite{BT98,HT12}. 
In order for fusion reactions to take place, the Coulomb barrier
between the colliding nuclei has to be overcome, and thus 
the fusion reactions occur by quantum tunneling at low energies. 
Moreover, atomic nuclei are composite particles, and their internal 
degrees of freedom can be excited during 
fusion. It has been well recognized by now that 
such excitations lead to a large enhancement of fusion cross sections 
at subbarrier energies as compared to a prediction of a simple one-dimensional 
potential model \cite{BT98,HT12,DHRS98,Back14}. 
One can consider this as a good example of coupling assisted tunneling 
phenomena. 

In order to analyze subbarrier fusion reactions,  
a coupled-channels approach has been recognized as 
a standard tool \cite{HT12}.  
In this approach, 
coupled-channels equations,  
with a few 
relevant excitation channels coupled strongly to the ground state, 
are numerically solved in order to 
obtain multi-channel penetrabilities 
\cite{HRK99}. 
In the eigen-channel representation of the coupled-channels 
approach, one diagonalizes the coupling matrix at each separation distance, 
obtaining 
$N$ different effective potential 
barriers, where $N$ is the number of channels to be 
included in the coupled-channels 
equations. 
The resultant multi-channel penetrabilities are then given as 
a weighted sum of the 
penetrability of each effective barrier \cite{DLW83,HB04}. 

Based on this idea, Rowley, Satchler, and Stelson have proposed a method 
to extract information on the distribution of fusion barriers 
directly from experimental 
fusion cross sections \cite{RSS91}. 
The idea is to take the second energy derivative of the quantity 
$E\sigma_{\rm fus}(E)$, where $E$ and $\sigma_{\rm fus}(E)$ are the incident 
energy in the center of mass frame and the fusion cross section, 
respectively, that is, $D_{\rm fus}(E)\equiv d^2(E\sigma_{\rm fus})/dE^2$. 
This quantity is referred to as fusion barrier distribution, and has played 
an important role in understanding the dynamics of 
subbarrier fusion reactions \cite{DHRS98,Leigh95}. 
For a single-channel problem, 
the quantity $D_{\rm fus}(E)$ is a Gaussian-like function 
centered at $E=B$, where $B$ is the height of the potential barrier. 
In multi-channel problems, the barrier distribution 
$D_{\rm fus}(E)$ has a multi-peaked structure, in which the position of 
each peak corresponds to the barrier height of each effective barrier.  
High precision measurements of fusion cross sections have been 
carried out, and the barrier distribution has 
been successfully extracted for many systems 
\cite{Leigh95}. It has been found that the 
barrier distribution is sensitive to the nature of channel couplings, 
revealing a fingerprint 
of multi-channel quantum tunneling \cite{DHRS98}. 
The concept of barrier distribution has been applied also to 
the dissociative adsorption of H$_2$ molecule in 
surface physics \cite{HT12}. 

In order to interpret the shape of fusion barrier distribution, 
the coupled-channels approach has been employed 
in most of the analyses \cite{HT12,DHRS98,DLW83}.  
That is, the experimental fusion barrier distributions are compared with 
theoretical distributions obtained with coupled-channels 
calculations for fusion cross sections 
$\sigma_{\rm fus}$. 
On the other hand, one may also consider 
a much simpler approach and fit directly the 
experimental fusion barrier distributions with a weighted sum of 
test functions in a single-channel problem. This simpler approach 
is yet helpful, as it 
may provide a useful feedback to quantal coupled-channels calculations, 
especially when the nature of couplings are not known {\it e.g.,} in 
exotic nuclei. For this purpose, one would have to 
optimize the number of test functions as well as 
parameters in the test 
functions. 
To this end, one may 
employ the chi-square fitting and minimize 
the chi-square function. 
However, this naive 
approach 
tends to lead to an overfitting problem, in which 
a data set is well fitted even when a model itself 
is inappropriate. For instance, 
$M$ data points can be fitted 
perfectly 
by introducing $M$ independent parameters, but such fit would be of no use. 

In this paper, I employ the Bayesian spectral deconvolution developed 
recently in the field of information science \cite{NSO12} 
in order to 
avoid 
such overfitting problem. 
In recent years, Bayesian approaches 
are becoming increasingly popular 
in nuclear physics, see {\it e.g.,} Refs. 
\cite{Papenbrock15,Furnstahl15,Wesolowski15,UPP16,SP16}. 
In the context of spectral deconvolution, an advantage of the Bayesian 
approach is that the number of peaks can be uniquely determined by 
introducing the stochastic complexity, with which the number of peak is 
determined by a balance between the chi-square and the complexity of the 
model. The aim of the present paper is to 
apply this approach to the fusion barrier distributions 
of $^{16}$O+$^{144,154}$Sm systems \cite{Leigh95} and discuss 
the usefulness of the method. 

Throughout this paper, I use the notation $P(A|B)$ to express the conditional probability of $A$ 
when $B$ is given. 
Suppose that one would like to fit a data set 
$D_{\rm exp}=\{E_i,d_i,\delta d_i\}$ ($i=1,2,\cdots M$), $\delta d_i$ being 
an experimental uncertainty of the quantity $d_i$, with 
a fitting function given by 
\begin{equation}
D_{\rm fit}(E;\tilde{\vec{\theta}},K)=\sum_{k=1}^K w_k\phi_k(E; \vec{\theta}_k). 
\label{eq:fit}
\end{equation}
Here, $\phi_k$ is a test function with a given parameter set 
$\vec{\theta}_k$, $K$ is the number of the test functions, and $w_k$ is 
a weight factor for each test function. 
The unknown parameters $\tilde{\vec{\theta}}\equiv\{w_k,\vec{\theta}_k\}$
with $k=1,2,\cdots K$ are determined by fitting to the data set, 
$D_{\rm exp}$. In the Bayesian approach, the experimental data $\{d_i\}$ are  
assumed to be realized as a sum of the fitting function 
given by Eq. (\ref{eq:fit}) 
and the uncertainty $\{\delta d_i\}$ \cite{NSO12}. 
That is, the conditional 
probability to find the data point $d_i$ for a given value of 
$\tilde{\vec{\theta}}$ and $K$ 
is given by 
\begin{equation}
P(d_i|E_i,\tilde{\vec{\theta}},K)=\frac{1}{\sqrt{2\pi (\delta d_i)^2}}\,
\exp\left(-
\frac{(d_i-D_{\rm fit}(E_i;\tilde{\vec{\theta}},K))^2}{2(\delta d_i)^2}
\right). 
\end{equation}
Given that the data points are realized independently to each other, the 
conditional probability 
for the whole $M$ data points thus reads  
\begin{equation}
P(D_{\rm exp}|\tilde{\vec{\theta}},K)=\prod_{i=1}^M 
P(d_i|E_i,\tilde{\vec{\theta}},K)
\propto e^{-\chi^2(\tilde{\vec{\theta}},K)/2},
\end{equation}
where 
$\chi^2(\tilde{\vec{\theta}},K)$ is the usual chi-square function given by 
\begin{equation}
\chi^2(\tilde{\vec{\theta}},K)
=\sum_{i=1}^M
\left(\frac{d_i-D_{\rm fit}(E_i;\tilde{\vec{\theta}},K)}{\delta d_i}\right)^2. 
\label{eq:chi2}
\end{equation}
Notice that the inclusive 
probability $P(D_{\rm exp}|K)$ for a given 
value of $K$ 
is calculated 
by integrating over all $\tilde{\vec{\theta}}$ as,
\begin{equation}
P(D_{\rm exp}|K)=\int d\tilde{\vec{\theta}}\,
P(D_{\rm exp}|\tilde{\vec{\theta}},K)P(\tilde{\vec{\theta}}),
\end{equation}
where I have assumed that the conditional probability 
$P(\tilde{\vec{\theta}}|K)$ is independent of $K$. 
Using the Bayes theorem, the conditional probability of $K$ for a given data set then reads,
\begin{equation}
P(K|D_{\rm exp})=\frac{P(D_{\rm exp}|K)P(K)}{P(D_{\rm exp})} \propto P(D_{\rm exp}|K),
\label{post}
\end{equation}
where I have assumed that the probability $P(K)$ is independent of $K$. 
Therefore, the most probable value of $K$ can be found by maximizing the function $Z(K)$ 
given by 
\begin{equation}
Z(K)=
\int d\tilde{\vec{\theta}}\,
e^{-\chi^2(\tilde{\vec{\theta}},K)/2}
P(\tilde{\vec{\theta}}),
\label{eq:z0}
\end{equation}
for a given 
prior probability, $P(\tilde{\vec{\theta}})$, or equivalently by minimizing the 
stochastic complexity (or the \lq\lq free energy\rq\rq) defined by 
$F(K)\equiv -\ln Z(K)$ \cite{NSO12}. 
Once the value of $K$ is so determined, the most probable value of the parameter $\tilde{\vec{\theta}}$ may 
be determined by minimizing the $\chi^2$ function given by Eq. (\ref{eq:chi2}). 

In order to evaluate the high dimensional integral in Eq. (\ref{eq:z0}), the authors of 
Ref. \cite{NSO12} have used the exchange Monte Carlo method \cite{Hukushima96}. 
To this end, they considered an auxiliary function given by 
\begin{equation}
z(\beta,K)=
\int d\tilde{\vec{\theta}}\,
e^{-\beta\chi^2(\tilde{\vec{\theta}},K)/2}
P(\tilde{\vec{\theta}}),
\end{equation}
and wrote the function $Z(K)$ in a form of 
\begin{equation}
Z(K)=\prod_{l=1}^{L-1}\frac{z(\beta_{l+1},K)}{z(\beta_l,K)}, 
\label{eq:z}
\end{equation}
with $\beta_1 < \beta_2 < \cdots < \beta_L$ (with $\beta_1=0$ and $\beta_L=1$). 
Notice that the function $z(\beta_{l+1},K)/z(\beta_l,K)$ in Eq. (\ref{eq:z}) is given by 
\begin{equation}
\frac{z(\beta_{l+1},K)}{z(\beta_l,K)}=
\frac{\int d\tilde{\vec{\theta}}\,e^{-(\beta_{l+1}-\beta_l)\chi^2(\tilde{\vec{\theta}},K)/2}q(\tilde{\vec{\theta}},K;\beta_l)}
{\int d\tilde{\vec{\theta}}\,q(\tilde{\vec{\theta}},K;\beta_l)}, 
\end{equation}
with the function 
$q(\tilde{\vec{\theta}},K;\beta_l)=e^{-\beta_l\chi^2(\tilde{\vec{\theta}},K)/2}
P(\tilde{\vec{\theta}})$, that is, the average 
of 
$e^{-(\beta_{l+1}-\beta_l)\chi^2(\tilde{\vec{\theta}},K)/2}$ 
with respect to $q(\tilde{\vec{\theta}},K;\beta_l)$. 
In the exchange Monte Carlo method, 
a set of the parameters $\tilde{\vec{\theta}}=\tilde{\vec{\theta}}_\beta$  
is generated independently 
for each value of $\beta$ according to the probability distribution 
$q(\tilde{\vec{\theta}},K;\beta)$ 
using {\it e.g.,} the Metropolis 
algorithm. That is, starting from an initial set of the parameters,  
the parameters at the $(n-1)$-th step, $\tilde{\vec{\theta}}_\beta(n-1)$, 
are successively updated for each $\beta$ 
to $\tilde{\vec{\theta}}_\beta(n)$. 
The parameters $\tilde{\vec{\theta}}_{\beta_k}(n)$ and $\tilde{\vec{\theta}}_{\beta_{k+1}}(n)$ (with 
$k=1,3,5,\cdots$ for even values of $n$ 
and $k=2,4,6,\cdots$ for odd values of $n$) 
are then exchanged (that is, $\tilde{\vec{\theta}}_{\beta_k}(n)\to\tilde{\vec{\theta}}_{\beta_{k+1}}(n)$ 
and $\tilde{\vec{\theta}}_{\beta_{k+1}}(n)\to\tilde{\vec{\theta}}_{\beta_{k}}(n)$) 
according to the probability of $v=\min(1,u)$ with 
\begin{equation}
u=\exp\left((\beta_{k+1}-\beta_k)[\chi^2(\tilde{\vec{\theta}}_{\beta_{k+1}}(n))
-\chi^2(\tilde{\vec{\theta}}_{\beta_{k}}(n))]/2\right).
\end{equation}
With the exchange procedure, local minima in the parameter space 
can be avoided to a large extent \cite{NSO12,Hukushima96}, and thus the initial value dependence 
for the Metropolis algorithm 
becomes considerably marginal. 
The optimum value and the uncertainty of $\tilde{\vec{\theta}}$ 
can be estimated {\it e.g.,} by 
taking the average and the variance of 
the samples $\tilde{\vec{\vec{\theta}}}(1), 
\tilde{\vec{\vec{\theta}}}(2), \cdots$ encountered during the 
Metropolis updates. 

\begin{figure}[t]
\includegraphics[clip,width=7.5cm]{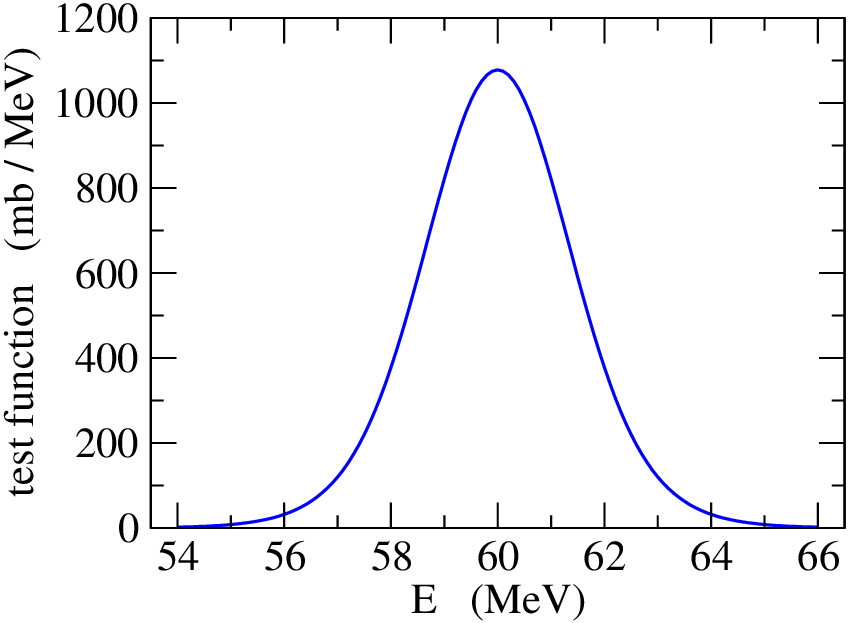}
\caption{(Color online) 
The test function defined by Eq. (\ref{eq:pdifference}). 
It is obtained with $B$ = 60 MeV, $R$ = 11 fm, $\hbar\Omega$ = 4.5 MeV, 
and $\Delta E$ = 1.8 MeV. }
\end{figure}

In order to apply the Bayesian spectral deconvolution to the fusion barrier 
distributions defined as $d^2(E\sigma_{\rm fus})/dE^2$, I 
employ a test function given by, 
\begin{eqnarray}
\phi_k(E;B_k,R_k,\Omega_k)&=&
\frac{1}{(\Delta E)^2}\,
[(E\sigma_{\rm fus})_{E+\Delta E}-2(E\sigma_{\rm fus})_{E} \nonumber \\
&& +(E\sigma_{\rm fus})_{E-\Delta E}],
\label{eq:pdifference}
\end{eqnarray}
where the fusion cross sections $\sigma_{\rm fus}(E)$ is given as, 
\begin{equation}
\sigma_{\rm fus}(E)=\frac{\hbar\Omega_k}{2E}R_k^2\,
\ln\left[1+\exp\left(\frac{2\pi}{\hbar\Omega_k}(E-B_k)\right)\right],
\label{eq:wong}
\end{equation}
using the Wong formula \cite{Wong73,RH15}. 
In Eq. (\ref{eq:pdifference}), the second derivative is evaluated with a 
point difference formula with the energy step $\Delta E$, as has been 
done in the experimental analyses \cite{DHRS98,Leigh95}. 
An example of the test function is shown in Fig. 1, which is obtained with 
$B$ = 60 MeV, $R$ = 11 fm, $\hbar\Omega$ = 4.5 MeV, 
and $\Delta E$ = 1.8 MeV. One can see that the test function is centered 
at $E=B$ with the full-width at half maximum of the order of 
0.5$\hbar\Omega$ \cite{RSS91}. 
Notice that with Eqs. (\ref{eq:fit}), (\ref{eq:pdifference}) and (\ref{eq:wong}), 
the independent parameters in this analysis read
$\tilde{\vec{\theta}}=\{\tilde{\vec{w}},\vec{B},\vec{R}\}$, where 
$\tilde{w}_k$ is defined as $\tilde{w}_k\equiv w_kR_k^2$. 

In the analyses given below, I take the prior probability distribution, 
$P(\tilde{\vec{\theta}})$, in Eq. (\ref{eq:z0}) as a uniform distribution 
in the rang of 
$54\leq B_k\leq 70$ MeV for $^{16}$O+$^{144}$Sm 
and 
$50\leq B_k\leq 65$ MeV for $^{16}$O+$^{154}$Sm, together with 
$0\leq \tilde{w}_k \leq 100$ fm$^2$ and 
$2 \leq \hbar\Omega_k \leq 5$ MeV (that is, 
$P(\tilde{\vec{\theta}})=0$ if
$\tilde{\vec{\theta}}$ is outside these ranges). 
Following Ref. \cite{NSO12}, I take $\beta_l=1.5^{l-L}$ with $L=24$ 
except for $l=1$, for which $\beta_1$ is taken to be zero. I also follow 
Ref. \cite{NSO12} to 
take 20000 points in the Metropolis algorithm for each $\beta_l$, for which 
the first 10000 points are thrown away as the burn-in period. 
For the initial values of the parameters $\tilde{\vec{\theta}}$, I take 
random values according to the probability distribution 
$P(\tilde{\vec{\theta}})$. 
I find that the 
results are sometimes 
improved if this procedure is iterated a few times, that is, 
if the Metropolis random walk is performed again starting from 
the optimum values of the parameters obtained in the previous 
Monte Carlo integration. 

\begin{figure}[t]
\includegraphics[clip,width=7.5cm]{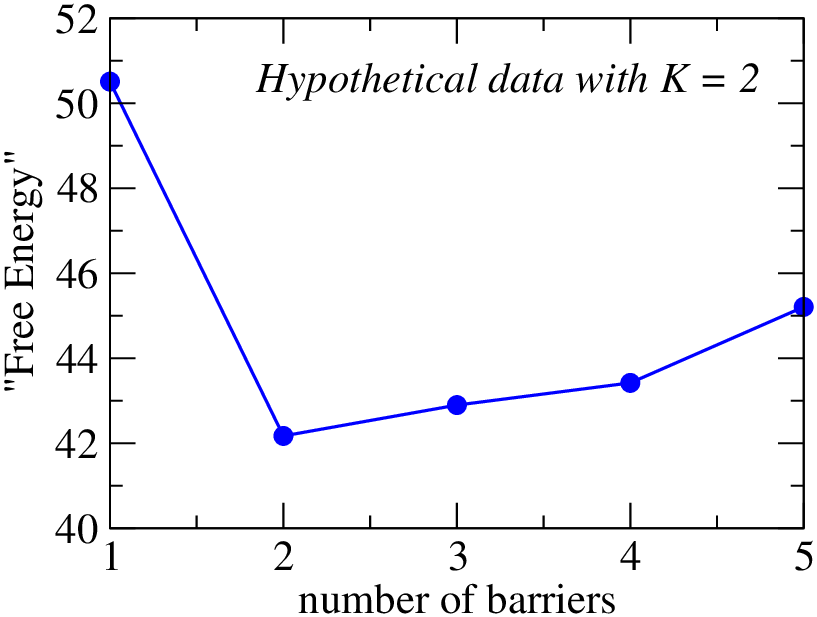}
\caption{(Color online) 
The ``free energy'', $F(K)=-\ln Z(K)$, as a function of the number of 
effective barrier, $K$, for hypothetical data generated with $K=2$. }
\end{figure}

Before I apply this procedure to the actual experimental data, I first 
carry out a proof-of-principle study by appling the method 
to hypothetical data generated with Eqs. (\ref{eq:fit}) and 
(\ref{eq:pdifference}) with $K=2$, 
$(\tilde{w}_1,B_1,\hbar\Omega_1)$ = (84.7, 60.0, 4.7), 
and $(\tilde{w}_2,B_2,\hbar\Omega_2)$ = (33.1, 65.0, 4.2), where 
$B$ and $\hbar\Omega$ are given in units of MeV while $\tilde{w}$ 
is given in units of fm$^2$. 
To this end, I generate a data set from $E$=55 MeV to 72 MeV with a step of 0.6 MeV, 
by adding randomly generated 1\% uncertainty in fusion cross sections, which 
is a typical value in high precision data \cite{DHRS98,Leigh95}. 
For the point difference formula for fusion barrier distribution, I take 
$\Delta E$ = 1.8 MeV, which is the value taken in the experimental analysis 
for the $^{16}$O+$^{144}$Sm system \cite{DHRS98,Leigh95}. 
As a result of the Bayesian spectral decomposition for this hypothetical data 
set, I find that $K$=2 indeed provides the minimum value of ``free energy'', 
$F(K)$, even though $K$=3 also gives a comparable fit (see Fig. 2). 
The optimum values for the parameters are found to be 
$(\tilde{w}_1,B_1,\hbar\Omega_1)$ 
= (84.4$\pm$1.91, 60.0$\pm$0.0301, 
4.70$\pm$0.0270), 
and $(\tilde{w}_2,B_2,\hbar\Omega_2)$ 
= (25.0$\pm$6.52, 64.5$\pm$0.600, 4.09$\pm$0.338). 
These values are consistent with the exact values, even though the agreement 
for the higher peak at $B=65$ MeV is less satisfactory due to larger error 
bars in the data (notice that the error bar in the barrier distribution 
increases as a function of energy, see also Fig. 3(c) below). 
Evidently, 
the Bayesian spectral deconvolution works well for an analysis 
of fusion barrier distributions. 

\begin{figure}[t]
\includegraphics[clip,width=7cm]{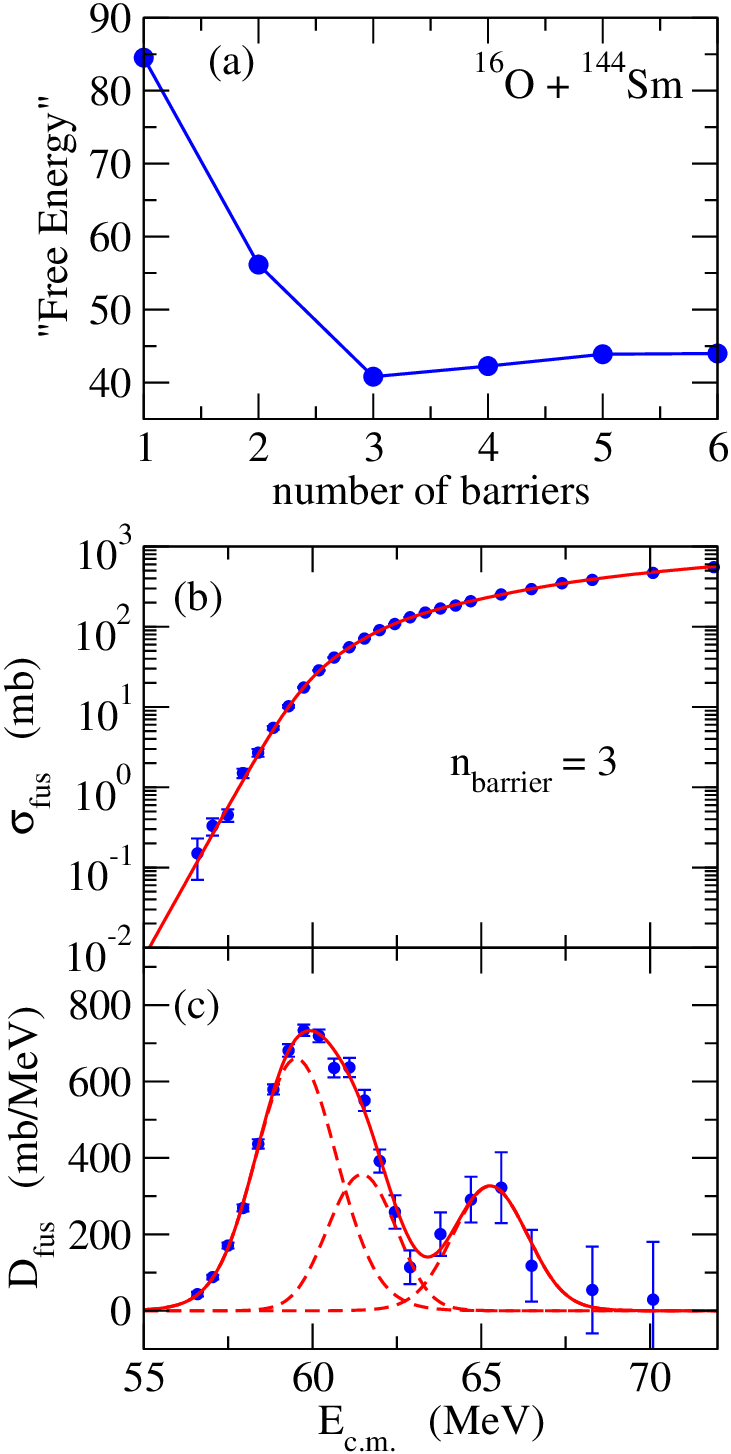}
\caption{(Color online) 
Top panel: the ``free energy'', $F(K)=-\ln Z(K)$, 
as a function of the number of effective 
barrier, $K$, for the $^{16}$O+$^{144}$Sm system. 
Middle panel: the fusion cross sections for the same system. 
The solid line is obtained with the 
Bayesian spectral deconvolution with $K$ = 3. 
Bottom panel: the corresponding 
barrier distribution defined 
by $d^2(E\sigma_{\rm fus})/dE^2$. 
The contribution of each test function is also 
shown by the dashed lines. 
The experimental data are taken from Ref. \cite{Leigh95}. 
}
\end{figure}

\begin{figure}[t]
\includegraphics[clip,width=7cm]{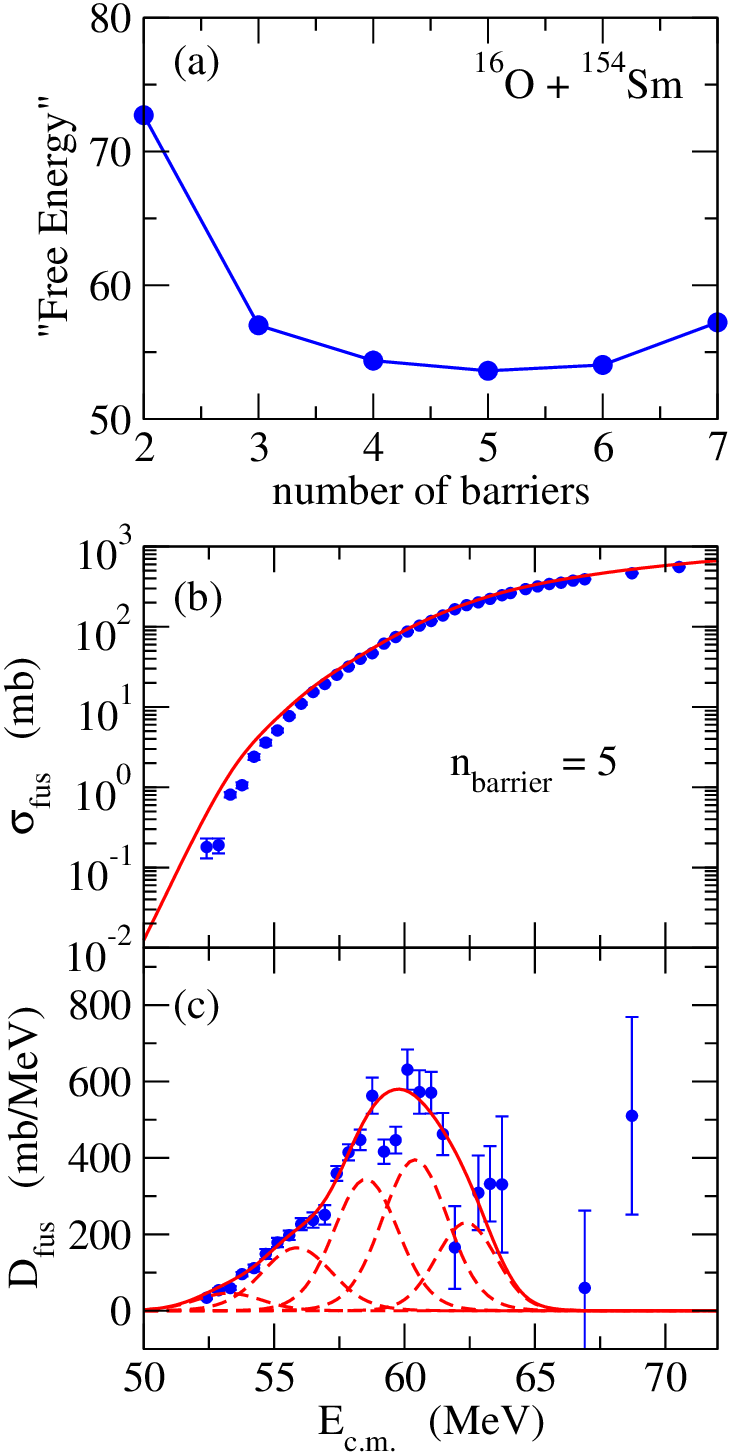}
\caption{(Color online) 
Same as Fig. 3, but for 
the $^{16}$O+$^{154}$Sm system. 
The middle and the bottom panels are obtained with $K=5$. 
}
\end{figure}

The posteriori probability of $K$, that is, 
$P(K|D_{\rm exp}) \propto e^{-F(K)}$ (see Eq. (\ref{post})), 
is evaluated as 
$P(K|D_{\rm exp})$ = 1.32$\times 10^{-4}$, 0.550, 0.266, 0.158, 
and 0.0264 for $K$ = 1, 2, 3, 4, and 5, respectively. 
In this paper, $K$=2 has been adopted among those 
as the optimum value of $K$ 
according to the idea of maximum-a-posteriori estimate \cite{NSO12}. 
As in Ref. \cite{NSO12}, I have repeated the same analysis by 
generating 100 independent data sets. I have found that  
the correct value of $K$, that is, $K=2$, is selected 78 times out of 100  
while $K$=3 and 4 were selected 19 and 3 times, respectively. 
The frequency to select the correct value of $K$ (that is, 78\%) 
is significantly larger than the posteriori probability 
$P(K=2|D_{\rm exp})=55.0 \%$,  
indicating the powerfulness of the present procedure \cite{NSO12}. 

Let us now analyze the experimental fusion barrier distribution for the 
$^{16}$O+$^{144}$Sm and the $^{16}$O+$^{154}$Sm systems. 
For the latter system, 
the data points at 
$E_{\rm c.m.}$ = 59.21 and 59.66 MeV are excluded 
in the fitting procedure, which appear to deviate from a 
smooth behavior of barrier distribution. 
The top panel of Figs. 3 and 4 shows the ``free energy'' as a function of 
the number of test function, $K$, thus the number of effective barrier, 
for the $^{16}$O+$^{144}$Sm and the $^{16}$O+$^{154}$Sm systems, respectively. 
It appears that the most probable value of $K$ is $K$ = 3 and 5 
for the $^{16}$O+$^{144}$Sm and the $^{16}$O+$^{154}$Sm systems, respectively. 
The optimum values of the parameters 
for these values of $K$ 
are summarized in Table I (in evaluating the average and the variance 
for each parameter for the 
$^{16}$O+$^{154}$Sm systems, I have increased the number of Metropolis 
sampling to 30000 in order to reduce the statistical error). 
The middle and bottom panels of Figs. 3 and 4 show the 
fusion cross sections and the fusion barrier 
distributions obtained with the optimum values of the parameters, respectively. 
The contribution of each test function 
is also shown by the dashed lines in the bottom panels. 
For the $^{16}$O+$^{144}$Sm system shown in Fig. 3, 
one can confirm that the experimental data are well 
fitted with this procedure. 
On the other hand, 
for the $^{16}$O+$^{154}$Sm system shown in Fig. 4, 
the fusion cross sections themselves significantly 
deviate from the experimental data at 
low energies, even though the fusion barrier distributrion is well reproduced. 
This is because the height of the 
lowest barrier has a large uncertainty (see Table I), to which the 
fusion cross sections are sensitive at energies well below the Coulomb 
barrier. 

\begin{table}[bt]
  \caption{
The optimum values for the fitting parameters obtained with the Bayesian 
spectral deconvolution. $K$ is the number of barrier and $w_k$ is 
the weight factor for each barrier. $B_k$, $R_k$, and $\hbar\Omega_k$ 
are the height, the position, and the curvature of each barrier, 
respectively. 
}
\begin{center}
\begin{tabular}{c|c|ccc}
\hline
\hline
System & $K$ & $B_k$ (MeV) & $w_kR_k^2$ (fm$^2$) & $\hbar\Omega_k$ (MeV) \\
\hline
$^{16}$O+$^{144}$Sm & 3 & 59.5$\pm$0.0789 & 64.3$\pm$6.06 & 3.58$\pm$0.149 \\
                  &   & 61.5$\pm$0.153 & 28.6$\pm$6.53 & 2.34$\pm$0.506 \\
                  &   & 65.3$\pm$0.251 & 29.0$\pm$4.63 & 3.00$\pm$0.338 \\
\hline 
$^{16}$O+$^{154}$Sm & 5 & 53.3$\pm$1.10 & 4.76$\pm$2.87 & 3.96$\pm$0.589 \\
                  &   & 55.9$\pm$0.467 & 18.3$\pm$3.16 & 4.40$\pm$0.438 \\
                  &   & 58.5$\pm$0.758 & 34.7$\pm$19.4 & 3.77$\pm$0.786 \\
                  &   & 60.4$\pm$0.833 & 40.7$\pm$21.4 & 3.90$\pm$0.939 \\
                  &   & 62.4$\pm$0.751 & 21.4$\pm$10.7 & 3.25$\pm$0.866 \\
\hline
\hline
\end{tabular}
\end{center}
\end{table}

It is interesting to notice that, for the $^{16}$O+$^{144}$Sm 
system shown in Fig. 3, the experimental fusion barrier distribution is 
consistent with the existence of three effective barriers, even though 
the experimental data show a clear double-peaked structure. I have confirmed 
that this conclusion remains the same even if the data point at $E$ = 60.64 
MeV is excluded from the fitting. 
For this system, the two main peaks in the barrier distribution has been 
interpreted to arise from the strong octupole vibrational excitation 
in the $^{144}$Sm nucleus \cite{Leigh95}. 
The existence of the third peak in the barrier distribution 
is consistent with the double-octupole 
phonon excitations
in the $^{144}$Sm nucleus, for which the effect of anharmonicity 
plays an important role \cite{HTK97}. 
On the other hand, 
for the 
the $^{16}$O+$^{154}$Sm system shown in Fig. 4, the barrier 
distribution obtained with the Baysian analysis is consistent with the 
distribution for a prolately deformed nucleus \cite{HT12}. 
If all the effective barriers arise from the rotational excitations of the 
$^{154}$Sm nucleus, $K$ = 5 indicates that excitations up to the 8$^+$ 
state plays an important role in the subbarrier fusion of the 
$^{16}$O+$^{154}$Sm system. 

In summary, I have applied the Bayesian spectral deconvolution 
in order to analyze the fusion barrier distributions for the 
$^{16}$O+$^{144,154}$Sm systems. 
Unlike a naive chi-square minimization, this approach 
provides a consistent way to 
determine 
the number of effective barriers originated from the 
channel coupling effects. I have successfully extracted the number 
of barriers from the experimental data. In particular, I have found that 
the barrier distribution for the 
$^{16}$O+$^{144}$Sm system is consistent with three barriers, which indicates 
the existence of double octupole vibration in 
the $^{144}$Sm nucleus. For the 
$^{16}$O+$^{154}$Sm system, on the other hand, I have confirmed the rotational 
nature of the target excitations. 
In this way, I have demonstrated the effectiveness of the Bayesian spectral 
deconvolution for fusion barrier distributions. 

In principle, one can 
perform similar studies also for barrier distributions defined with 
heavy-ion quasi-elastic scattering at backward angles \cite{Timmers95,HR04}. 
It would be an interesting future study to apply the Bayesian spectral 
deconvolution both to the fusion and to the quasi-elastic barrier 
distributions for the same systems and discuss the consistency of the 
barrier parameters in the different reaction processes. 

\medskip

I thank T. Ichikawa for useful discussions and for his careful reading of the manuscript.

\end{document}